\documentstyle[psfig]{mn}
\def\refitem{\par\parskip 0pt\noindent\hangindent 20pt}
\def\etal{{et\thinspace al.}\ }

\title[Unifying models for BL Lac Objects]
{Unifying models for X--ray selected and Radio selected BL Lac Objects}
 
\author[G. Fossati, A. Celotti,  G. Ghisellini, L. Maraschi]
{G. Fossati$^1$, A. Celotti$^1$, G. Ghisellini$^2$, L. Maraschi$^{3}$ \\ 
$^1$ SISSA/ISAS, via Beirut 2--4, 34014 Trieste, Italy \\ 
$^2$ Osservatorio Astronomico di Brera--Merate, Via Bianchi 46, I-22055 Merate (Lecco), Italy \\
$^3$ Osservatorio Astronomico di Brera, via Brera 28, 20121 Milano, Italy\\}

\date{Received ***; in original form ***} 
       
\begin{document} 

\maketitle 

\begin{abstract} 

We discuss alternative interpretations of the differences in the Spectral
Energy Distributions (SEDs) of BL Lacs found in complete Radio or X--ray
surveys.

A large body of observations in different bands suggests that the SEDs of
BL Lac objects appearing in X--ray surveys differ from those appearing in
radio surveys mainly in having a (synchrotron) spectral cut--off (or
break) at much higher frequency. 

In order to explain the different properties of radio and X-ray selected
BL Lacs Giommi and Padovani proposed a model based on a common radio
luminosity function. At each radio luminosity, objects with high frequency
spectral cut-offs are assumed to be a minority. Nevertheless they dominate
the X-ray selected population due to the larger X--ray--to--radio--flux 
ratio. 
An alternative model explored here (reminiscent of the orientation models
previously proposed) is that the X-ray luminosity function is ``primary''  
and that at each X-ray luminosity a minority of objects has larger
radio--to--X--ray flux ratio.

The predictions of the two scenarios, computed via a Montecarlo technique,
are compared with the observed properties of BL Lacs in the two samples
extracted respectively from the 1 Jy radio survey and the {\sl Einstein}
Slew Survey. We show that both models can explain a number but not all the
observed features.

We then propose a completely new approach, based on the idea that the
physical parameter which governs the shape of the SEDs, is (or is
associated with) the bolometric luminosity. Assuming an empirical relation
between spectral shape and luminosity we show that the observational
properties of the two surveys can be reproduced at least with the same 
accuracy as the two previous models. 

\end{abstract} 


\section{Introduction}

Among Active Galactic Nuclei, Blazars show extreme luminosity and
variability. The observed properties are best interpreted in the frame of
the relativistic jet model, proposed almost twenty years ago by Blandford
\& Rees (1978), as due to relativistic ``beaming'', i.e. the effects of the
relativistic motion of the emitting plasma on the observed radiation.  In
particular BL Lac objects, which are considered part of this class, are
characterized by their almost featureless continua.

BL Lacs have been almost exclusively discovered through radio or X--ray
surveys. However the properties of objects selected in the two spectral
bands are systematically different, posing a question as to whether there
are two ``types'' of BL Lacs.  The first difference to be recognized and
perhaps still the most striking is the shape of the SED. The differences
show up using broad band spectral indices and color--color diagrams, e.g.
$\alpha_{RO}$ vs $\alpha_{OX}$\footnote{$\alpha_{1,2} \equiv -
\log(F_1/F_2)/\log(\nu_1/\nu_2)$; radio fluxes are taken at 5 GHz, optical
at 5500 \AA, X--ray at 1 keV} (Stocke \etal 1985; Maraschi \etal 1995;
Sambruna, Maraschi \& Urry 1996). Others include optical polarization,
variability,
presence of (weak) emission lines, radio luminosity, core--dominance, all
of which are less conspicuous in X-ray selected objects (e.g. Kollgaard
\etal 1992; Perlman \& Stocke 1993; Jannuzi \etal 1994; Kollgaard \etal
1996; see Kollgaard 1994 for a recent review).

On the basis of the X-ray to radio flux ratio (or $\alpha_{RX}$ value)  we
can define an objective criterion (independent of the selection band)
separating two (putative) classes of objects: we define XBLs the objects
with $\log (F_{\rm 1keV}/F_{\rm 5GHz}) \ge -5.5$ and RBLs the objects with a
ratio smaller than this dividing value, where the fluxes are monochromatic and
expressed in the same units (see also Wurtz 1994, Giommi \& Padovani 1994, 
for an analogous definition).
As can be seen from Fig.~1a XBLs are found
mostly but not exclusively in X-ray surveys and the same is true for RBLs
with respect to radio surveys. 

The spread in spectral shapes was originally attributed to orientation
effects, associated with different widths of the beaming cones of the
radio and X-ray radiation emitted by a relativistic jet (Stocke \etal
1985; Maraschi \etal 1986; Celotti \etal 1993). The idea came from the
observational evidence that BL Lacs discovered in radio and X--ray surveys
actually show similar X--ray luminosities while the radio luminosities
typically differ by two--three orders of magnitude (e.g.  Fig.~1b). This
could be accounted for if X--ray radiation had a wider beaming cone than
radio emission:  observers would see similar X--ray luminosities over a
wide range of angles while the accompanying radio luminosity would be high
for a small fraction of objects seen at very small angles and strongly
dimmed for the majority, observed at larger angles (e.g. Fig.~5 in Celotti
\etal 1993).  Consequently the number density ratio between the two
``flavours'' would be determined by the associated solid angles. Since the
X--ray emission is largely isotropic, X--ray surveys are not biased
against any of the two classes of objects, and can give the correct number
ratio. The different beaming affecting the various bands could be due to
an accelerating (Ghisellini \& Maraschi 1989) or an increasingly
collimated jet (Celotti \etal 1993).

Substantial progress in the data on the SEDs of BL Lac objects has been
obtained in recent years, in terms both of sensitivity and statistics,
especially in the X-ray band (Giommi, Ansari \& Micol 1995; Comastri,
Molendi \& Ghisellini 1995; Perlman \etal 1996a,b; Urry et al. 1996; 
Sambruna et al. 1996). 
In terms of the power emitted per decade XBLs
display a continuous rise up to the UV and in extreme cases the soft X-ray
band, while RBLs are characterized by a spectral turn over in the IR
domain. In XBLs the X-ray emission is dominated by a soft spectral
component which extrapolates continuously to lower frequencies.  In RBLs
the X--ray emission is dominated by a separate harder component which at
least in some cases extends to the $\gamma$--ray domain (von Montigny et al. 
1995; for X--ray spectra see the results by Padovani \& Giommi (1996) and 
Lamer, Brunner \& Staubert 1996).
A hard GeV to TeV component is present also in XBLs though it does
not show up in medium energy X-rays. Sambruna et al. (1996) showed that it
is difficult to model the detailed transition from an XBL to an RBL like
in terms of orientation only and suggested rather a continuous change in
the physical parameters of the jet.

The first model developed along this line to explain the ``statistics'' of
XBLs and RBLs is the ``radio luminosity + different energy cutoff''
scenario by Giommi and Padovani (Giommi \& Padovani, 1994; Padovani \&
Giommi 1995;  hereafter we refer to it as the ``radio leading'' scenario by
G\&P) alternative in many ways to the commonly accepted ``different
viewing angle'' model sketched above. They propose that a single
luminosity function in the radio band describes the full BL Lac
population. For each radio luminosity X--ray bright BL Lacs (i.e. XBLs) 
are intrinsically a minority described by a fixed (luminosity independent)
distribution of X--ray to radio flux ratios\footnote{We note that the
definition of XBL and RBL by G\&P is based on the X--ray flux in the
0.3$-$3.5 keV {\sl Einstein} IPC band (in erg/cm$^2$/sec) and the radio
flux at 5 GHz (in Jy). According to this different definition, the value
of $-$5.5 adopted here corresponds to $\sim -10.8$}. In X--ray surveys
however selection effects substantially enhance the XBL fraction.
According to this approach the intrinsic fraction of the two types of BL
Lac (XBL vs. RBL)  would be objectively reflected in radio surveys. With
these hypotheses G\&P were able to reproduce the observed
X--ray counts, luminosity functions and the distribution of BL Lacs in 
the $\alpha_{\rm RO}-\alpha_{\rm OX}$ plane.

Because of the implications of these issues for physical models of
relativistic jets and the understanding of the physical conditions within
the emission region, we decided to explore more thoroughly the fundamental
hypothesis that BL Lacs are a single class, whose SEDs are characterized
by different physical parameters, and test any prediction against the
observations now available.

In addition to the radio leading model we consider a first ``symmetric'' 
alternative, namely that the X-ray luminosity function basically
represents the whole BL Lac population.  In this case X--ray surveys would
give objective results regarding the intrinsic abundance of XBLs and RBLs.
This scenario, which we refer to as the ``X--ray leading'' model, can be
considered as an evolution of the ``different viewing angle'' scenario,
where the X--ray luminosity was the basic property and the population
ratio reflected the ratio of solid angles.

The predictions of both models, derived by a Montecarlo technique, are
compared in detail with the observations now available. A main feature of
the samples, that is the redshift distribution, appears poorly reproduced
by either models. 

We then propose a new, unified picture, in which the key feature is a link
between the shape of the SEDs, in particular the peak frequency of the
synchrotron power distribution, and {\it the bolometric luminosity}. In
particular, we parameterize the shape of the SED in terms of its
synchrotron peak frequency and assume a power law relation between peak
frequency and bolometric luminosity. Adopting an estimate of the
bolometric luminosity function we can derive all the observable properties
and again compare them with observations.

We test all the scenarios against the best (whole) body of data now
available:  the radio and X--ray fluxes (uniformly measured by {\it
ROSAT}, Urry \etal 1996) of the 1 Jy BL Lac sample (Stickel \etal 1991) 
and the radio and X--ray fluxes of the BL Lac sample derived from the {\sl
Einstein} Slew survey sample (Elvis \etal 1992; Perlman \etal 1996a).
Based on the data and the discussion by Perlman et\thinspace al., 
we treat the Slew survey sample available to date as a tentative, 
but quasi--complete one.

In Section 2 we present the radio and X--ray samples of BL Lacs against
which we tested the predictions of the models.  The two alternative
``radio and X--ray leading'' scenarios, as they will be called hereafter,
are outlined in Section 3, while we describe the simulation procedure
adopted for the comparison with the observed samples in Section 4. In
Section 5 we present and compare the predictions of the two models with
observations and briefly discuss the effects of evolution. The assumptions
and predictions of the new ``unified bolometric'' model are described in
Section 6. Section 7 summarizes our conclusions.

The values $H_0 = 50$ km/sec/Mpc and $q_0 = 0.0$ have been adopted
throughout the paper.

\section{The reference samples}

\subsection{The 1 Jy sample}

The complete 1 Jy BL Lac sample was derived from the catalog of radio
selected extragalactic sources with $F_{\rm 5GHz} \ge 1$ Jy (K\"uhr \etal
1981) with additional requirements on radio flatness ($\alpha_R \le 0.5$,
with $F_{\nu} \propto \nu^{-\alpha}$), optical brightness ($m_V \le 20$)
and the absence of optical emission lines (EW$_{\lambda} \le 5$ \AA,
evaluated in the source rest frame) (Stickel \etal 1991). This yielded 34
sources matching the criteria, 26 with a redshift determination and 4 with
a lower limit on it (Stickel \etal 1994). It is the largest complete radio
sample of BL Lacs compiled so far. 

We computed the luminosities using the monochromatic 1 keV fluxes
measured by {\it ROSAT} (Urry \etal 1996, where the listed values are
those from fits with Galactic absorption) and the 5 GHz values from the
K\"uhr \etal catalogue. We considered only the subsamples of sources
with at least a lower limit on the redshift. The fluxes were
K--corrected using a radio spectral index $\alpha_R = - 0.27$ and the
average X-ray spectral index measured in the {\it ROSAT} 
band, $\alpha_X = 1.16$. 

\subsection{The Slew survey sample}

The {\sl Einstein} Slew survey (Elvis \etal 1992) was derived from data
taken with the IPC in between pointed observations. A catalog of 809
objects has been assembled with a detection threshold fixed at 5 photons. 
It does not reach high sensitivity, having a flux limit of $\simeq 5
\times 10^{-12}$ erg/cm$^2$/sec in the IPC band (0.3 $-$ 3.5 keV), but it
covers a large fraction of the sky ($\sim$ 36600 deg$^2$).  Based on radio
imaging and spectroscopy Perlman \etal (1996a) selected from this sample a
set of 62 BL Lac objects (33 previously known and 29 new candidates) and,
in a restricted region of the sky, a quasi--complete sample of 48 BL Lacs. 
The sources are almost completely identified, and therefore constitute a
quasi--complete sample of 48 BL Lacs. The redshift is known for 41 out of
48 objects.  This is the largest available X--ray selected sample of BL
Lacs (Perlman \etal 1996a), with more than twice the number of sources
contained in the fairly rich X--ray selected sample derived from the {\sl
Einstein} Medium Sensitivity Survey (EMSS, Morris \etal 1991, Wolter \etal
1994;  Perlman \etal 1996b), extended to 23 objects by Wolter \etal
(1994). 

K--corrections of the radio and X--ray monochromatic fluxes were
computed using spectral indices $\alpha_R = 0.0$ and $\alpha_X = 1.5$,
respectively.

\section{The two scenarios}

\subsection{Different energy cut--off}

Giommi and Padovani introduced the idea that BL Lacs are a single
population of objects whose SED can be characterized phenomenologically by
the distribution of the values of the frequency at which the peak in the
energy emitted per logarithmic bandwidth occurs (i.e. the peak in the $\nu
F_{\nu}$ representation of the broad band energy distribution) for the
putative synchrotron component.

A good quantitative indicator of the overall broad band shape of SEDs is
the ratio between X--ray and radio fluxes, conventionally taken at 1 keV
and 5 GHz, equivalent to a characterization in terms of $\alpha_{RX}$ and
$\alpha_{RO}$ (for a representation of the relation between SED shape and
for instance $\alpha_{RO}$ see Fig.~2 in Maraschi \etal 1995;
Comastri et al. 1995 tested also the
$\alpha_{RX}-\nu_{\rm peak}$ correlation).  
SEDs peaking at lower frequencies correspond to higher ratios (e.g. Maraschi
et al. 1995).  
The X--ray to radio flux ratio is typically two orders of magnitude larger
in BL Lac objects derived from X--ray surveys than in objects derived from
radio surveys.

It is clearly more convenient to work with flux ratios since they are
easier to determine then the peak frequency of a broad band energy
distribution.  One can even observationally derive a distribution for the
X--ray/radio luminosity ratio as if this is the relevant intrinsic
quantity which characterizes the SED distribution. This is what we
consider hereafter, following the G\&P approach.

Starting from these hypothesis, one can derive the statistical properties
of BL Lacs simply by assuming a radio (X--ray) luminosity function and a
probability distribution, $\cal P$, of the X--ray to radio luminosity
ratio.  In the following two subsections we present the assumptions on
these two quantities, according to the radio leading model and its
symmetric, in which the X--ray luminosity is the leading one.

\subsection{``Radio luminosity leading''}

The basic radio leading (G\&P) hypothesis (Giommi \& Padovani 1994; 
Padovani \& Giommi 1995) is that radio selection would be objective with
respect to the intrinsic spread of broad band spectral properties. The
underlying idea is that the radio emission is only weakly affected by the
properties of the synchrotron component at higher energies, such as the
peak frequency. Therefore the radio selection is not expected to suffer of
any bias regarding the SED shape properties and sources with different
X--ray/radio luminosity ratios are sampled from a common radio luminosity
function.

In this approach, the X--ray counts (and X--ray luminosity function) are
easily predictable from the radio counts (and radio luminosity function) 
and the results can be compared with real data coming from surveys.  It is
worth reminding that there are essentially no free parameters in the
model.

Since a radio selected sample would be unbiased with respect to the X--ray
to radio luminosity ratio, G\&P considered the 1 Jy sample of BL Lacs
(Stickel \etal 1991) and built the probability distribution $\cal P$. 
To improve the statistics for the XBL--like objects part of $\cal P$, 
we considered, following Padovani and Giommi (1995), the distribution 
deduced from the EMSS sample, weighted by a factor 1/10. 
In fact the ratio of XBL to RBL objects derived
from various radio selected samples (1 Jy, Stickel \etal 1991;  S4,
Stickel \& K\"uhr 1994; S5, Stickel \&  K\"uhr 1996) covers the range
$N_{\rm XBL}/N_{\rm RBL}= $ 1/16 $-$ 1/6. Therefore, with the given
hypothesis,
this factor 1/10 represents the intrinsic ratio of XBL to RBL populations
in the whole BL Lac class: in the radio leading scenario only a minority
of sources would intrinsically have the SED peaking in the UV range, while
most of the energy distributions would peak in the $\sim$ IR range. 

To calculate the luminosity function in the luminosity range $10^{28 -
36}$ [erg/sec/Hz], G\&P extrapolated the observed one to lower
luminosities. This extrapolation is based on the beaming model scheme (see
Urry \etal 1991), applied to the luminosity function derived from the 1 Jy
sample data (covering the range $10^{32} - 10^{35}$ erg/sec/Hz, Stickel et
al. 1991), in the non evolutionary case.  We note that the predictions of
the model are indeed strongly dependent on the number density of the low
luminosity sources.

Clearly, the assumption of a minimum radio luminosity $L_{\rm 5GHz,min}$
introduces a limiting luminosity in the X--ray band, $L_{\rm 1keV,min}$. 
However, it is worth noting that, because of the different spectral shape
of XBL and RBL objects, their corresponding minimum luminosities will be
different, with $L_{\rm 1keV,min}|_{\rm RBL}< L_{\rm 1keV,min}|_{\rm 
XBL}$.

G\&P found a good agreement between the predicted X--ray counts and those
derived from all the main X--ray surveys (EMSS, EXOSAT High Galactic
Latitude Survey, HEAO-1 surveys; see Fig.~3 of Padovani \& Giommi 1995) 
and with the EMSS luminosity function. 

In their model, the observational evidence that XBL objects are more
numerous, apparently in contradiction with their assumption that XBL
sources are intrinsically only 1/10 of the entire population, is due to
selection effects. In fact, given their $F_{\rm 1keV}/F_{\rm 5GHz}$
ratios, XBL have a relatively low radio flux. Therefore, at the same
X--ray flux level we are detecting together BL Lacs belonging to different
parts of the radio luminosity function, with RBLs coming from a brighter
but poorer part of it.

Our aim here is to perform a further step in checking the radio leading
scenario, i.e. to compute the distribution of radio and X--ray
luminosities predicted for samples selected in the ``leading'' (radio) band
also in the ``secondary'' (X--ray) spectral band, using the same inputs of
G\&P. 

\subsection{``X--ray luminosity leading''}

The first alternative scenario that we tested is substantially
``symmetric'' to the radio leading one. It assumes that the common
property of XBL and RBL is the X--ray luminosity.  As already mentioned,
this idea originally comes from the observational evidence that BL Lacs
discovered in radio and X--ray surveys actually show similar X--ray
luminosities, while the radio luminosities typically differ by two--three
orders of magnitude (e.g. Maraschi \etal 1986; see also Fig.~1b). This is
in general true also if the distinction between the two types of sources
is made in terms of the RBL/XBL classification as formally defined in
Section 1 (see also Fig.~1a).

Again, the starting points are the X--ray luminosity function and a
probability distribution, $\cal P$, of luminosity ratios, but in this case
this should be observationally derived from an X--ray survey, that by
construction is now supposed to be the objective one. 

An important assumption, which we had necessarily to take into account in
the X--ray case, concerns the cosmological evolution of BL Lacs. While it
has been so far established that RBLs show a slight positive evolution,
consistent with no evolution at the 2$\sigma$ level (e.g. Stickel \etal 
1991; Wolter \etal 1994),
recently Wolter \etal (1994) and Perlman \etal (1996b)  confirmed the
strong negative evolution in the X--ray band: X--ray bright objects are much
less luminous or common at high redshifts. In analogy with the radio case, we
first evaluated the predictions of the X--ray leading model assuming no
evolution. The results have been presented by Fossati \etal (1996)
who showed that they are not compatible with the 1 Jy survey properties, 
predicting a number of RBL objects in large excess with respect to the
observed one.

Then, in order to be able to reproduce the observational results and in
particular the redshift distributions (see below), here we assumed a
negative luminosity evolution in the X--ray band.

We derived the distribution $\cal P$ from the Slew survey X--ray selected
sample, but using only objects with redshift lower than 0.25, in order to
better approximate a distribution appropriate for a sample of sources at
the same redshift. 

Unfortunately, the X--ray luminosity function for this sample is not (yet)
available.  For this reason we considered the EMSS and its well studied
X--ray luminosity function (Morris \etal 1991; Wolter \etal 1994). We
tentatively adopted an X--ray luminosity function matching that derived by
Wolter et al. (1994), represented as a single power law with slope
$\alpha_{\phi}$ = $-$1.62. Since the normalization of the luminosity function
would substantially affect only the total number of sources, but neither
their relative number nor the average luminosities, in order to recover
the correct absolute number of objects we fixed the normalization (for the
case with best results reported in Table~1) at the value $\log \Phi_0
(L_{\rm 1keV})= -36.3$. This value is within the confidence range
given by
Wolter \etal (1994), with $L_{\rm 1keV}$ corresponding to
$L_{\rm 0.3-3.5keV}
= 10^{40}$ erg/sec.

The luminosity evolution parameter is $\tau$ = 0.142, according to the
definition $L(z) = L(0) \exp(-t_{LB}/\tau)$, where $t_{LB}$ is the
look--back time in units of t$_0$ (for $q_0 = 0.0$, t$_0$ = 1/H$_0$). The
luminosity range is $10^{24 - 30}$ [erg/sec/Hz]. Also in this case we
stress that RBLs and XBL reach different values of the minimum radio
luminosity (see Section 3.2).

\section{The simulations} 

Given these assumptions, we computed the predictions of both models. We
used a Montecarlo technique to simulate the distribution of sources in
space and luminosity. This method is very convenient because it allows us
to store ``single source'' attributes and not only to compute sample
integrated average properties. In other words, we can actually simulate a
catalogue and from it compute the desired average quantities.

For the ``radio leading'' scenario we used a third degree polynomial fit
to the luminosity function published in Fig.~1 of Padovani \& Giommi
(1995)\footnote{The polynomial fit is expressed as $\log \Phi = a + b
\log L_{\rm 5GHz, 28} +c (\log L_{\rm 5GHz, 28})^2 + d (\log L_{\rm  
5GHz, 28})^3$ Gpc$^{-3}$ L$^{-1}$, where $L_{\rm 5GHz, 28} = 10^{-28}
L_{\rm 5GHz}$ erg/s/Hz, and $a= -25.204616$, $b= - 2.1376271$, $c= -
0.0455836$ and $d= - 0.0059641$}. In all cases simulations have been
performed within the redshift range $0 - 2$.

In Fig.~2a,b the observed $L_{\rm 1keV}/L_{\rm 5GHz}$ distributions are
shown, taking into account only objects with redshift estimate. The dashed
histograms represent RBL objects, the dotted ones XBLs.  Note that there
is a superposition of the two classes due the fact that the actual
distinction between RBL and XBL is based on the flux ratio (while here the
luminosities are considered). In Fig.~2c the distribution $\cal P$, used
in the calculations for the radio leading scenario, is shown.

The actual computation can be summarized in five steps: 

\begin{enumerate} 

\item ``draw'' a redshift and a luminosity from the
luminosity function assumed as primary (radio for the radio leading,
X--ray for the X--ray leading scenarios);

\item the ``secondary'' luminosity (i.e. in the other spectral band) is
deduced by choosing a luminosity ratio by means of its probability
distribution $\cal P$;

\item radio and X--ray fluxes are computed for the simulated source taking
into account the ``inverse'' K--correction\footnote{The ``inverse'' refers
to the fact that usually the K--correction is meant to convert
observer--frame to source--frame quantities, while here we are using it in
the other direction.};

\item check if the simulated source would be detected in a survey with a
given flux limit and, if so, store its set of parameters. 

The detectability criteria take into account the sky coverage $\Delta
\Omega$ that in the radio case is a step function of the monochromatic 5
GHz flux, while in the X--ray survey it is an increasing function of the
flux integrated over the whole IPC band (0.3 $-$ 3.5 keV). In the X--ray
case we then simulated N surveys (typically 15 $-$ 20) each with a
different flux limit $F_i$ and a corresponding $\Delta \Omega_i$. The
results of the N surveys are then summed up to construct the full X--ray
survey. 

The flux dependent sky coverage for the Slew survey has been derived from
the diagram ``exposure time'' vs. ``percentage of sky'' published in Elvis
\etal (1992). A limiting flux corresponding to a given exposure time is
deduced considering a detection limit of 5 IPC counts, and a conversion
factor from IPC counts to flux of $3.26 \times 10^{-11}$ erg/cm$^2$/sec
(Elvis \etal 1992), valid for values of $\alpha_X = 1.2$ and
$N_H = 3 \times 10^{20}$ cm$^{-2}$. 
In fact, in order to convert to monochromatic flux we adopted an X--ray 
spectral index $\alpha_X = 1.2$, a fiducial average value for the class of 
sources that we are considering. This gives the conversion relation:  
$\log F_{\rm 1keV} = \log F_{\rm [0.3-3.5keV]} - 17.776$. 
It is important to stress here that we use an approximated sky coverage 
by using the information currently available. 
An exact correction cannot be computed, but plausibly this would mostly 
affect the {\it normalization} of the assumed luminosity functions.

\item finally, a detected source is classified as XBL or RBL according
to its $F_{\rm 1keV}/F_{\rm 5GHz}$ ratio, as already defined. 

\end{enumerate}

\section{Results and comparison of the models} 

The comparison of the two models with the 1 Jy and Slew survey samples are
presented in Figs.~3 and 4, respectively.  The average observed and
predicted radio and X-ray luminosities are reported in Table~1, together
with the numbers of objects. 
The consistency of the predicted and observed distributions has been 
quantitatively estimated through the use of the Student's t--test
for comparison of the means, and also of the more sophisticated 
survival analysis non-parametric and univariate methods (Feigelson \&
Nelson 1985) allowing to take into account lower limits. 
Student's t--test has been performed on all average luminosities, while 
the survival analysis has been applied only to the cases with a
significant number of objects, namely for the RBLs subsample in the 
``1 Jy'' and XBLs in the Slew survey.
The results of the statistical tests are reported as an entry in Table~1 
for Student's t--test, and separately in Table~2 for the survival analysis.

\smallskip
\noindent {$\bullet$}{\it Radio survey} 

\noindent
Let us consider first the comparison of the models with the 1 Jy sample
(Fig.~3). 

Both models predict luminosities for the XBL in both bands more than one
order of magnitude larger than observed. However the extreme paucity of 
objects does not allow these discrepancies to be significant from a 
statistical point of view. The Student's t--test indicates
that the radio leading scenario predicts a disagreement, at the $>$ 94.0 per 
cent level, in the X--ray luminosity of the RBL, while the 
X--ray leading one has difficulties at reproducing the average luminosity 
of the XBL at the 91 per cent level. 

The relative number of objects detected in both scenarios is in reasonably
good agreement with the observed one (5/29 and 3/27 instead of 2/32).

\smallskip
\noindent {$\bullet$}{\it X--ray survey} 

\noindent
The predictions of the two models concerning the Slew survey objects are
instead quite different (Fig.~4). 

The radio leading scheme leads to incorrect X--ray and radio luminosities
of XBL: from survival analysis 97.2 and 99.96 per cent levels (Table~2), 
respectively, even worst according to the Student's t--test (Table~1).  
In particular, in the radio leading scenario the average radio luminosities 
of the RBL and XBL subsamples tends unavoidably to be the same:  
as immediate consequence of the ``radio leading''
hypothesis, all BL Lac objects should share the same range of radio
luminosity, in contrast with observations of complete samples (see also
Fig.~1b). The average radio luminosity of XBL is also overestimated within
the X--ray leading scenario ($> 95.6$ per cent level). 
Again, nothing can be said from these tests on the distributions of the 
less numerous source population. 

The predicted numbers of XBLs and RBLs in the Slew survey (43/7 and 43/7)
are in good agreement with the observed one (40/8). 

\vskip 0.2 truecm
\noindent 
Summarizing, in terms of source numbers and average luminosities
the X--ray leading model is better than the radio leading one, sharing with
it the problem of XBL luminosities in radio surveys.  Nevertheless, it
should be noted that XBL objects in the 1 Jy survey are only 2 and so
their measured properties could mis--represent the true ones.

\smallskip
\noindent {$\bullet$}{\it Redshift distribution} 

\noindent
Finally, there is a significant disagreement on the red-shift distributions
(see the panels of the first columns in Figs.~3 and 4): both models
predict a sharp peak of simulated RBL sources at low redshift, in
contrast with the observed flat distributions\footnote{Note that the
inconsistency between the distribution predicted by the radio leading
scenario and the one observed in the 1 Jy sample is 
in agreement with the assumption of no evolution adopted here. 
If we had introduced the slight positive evolution apparent from ``1 Jy'' 
data we would have obtained a redshift distribution completely consistent 
with the observed one.}.
  
This is a {\it built--in} feature of this kind of approach in which we
take the relative number of the the two types of sources fixed for any
value of the leading luminosity.

Note however that, as expected, the inclusion of the negative evolution in
the X--ray leading model tends to improve the agreement with
observations.
Indeed the redshift distributions obtained in the X--ray
leading non evolving case were qualitatively similar to the ones of the
radio leading scenario (see Fossati et al. 1996). 
In fact in the 1 Jy
survey XBLs are sharply concentrated at lower redshifts, while RBLs show a
``flatter'' distribution. These trends are closer to the observed ones than
those derived in the radio leading scenario (Fig.~3). The same is true for the
Slew survey (Fig.~4). 

The use of the survival analysis allowed us to perform 
a quantitative analysis. 
The comparison of the only RBLs in the 1 Jy survey
indicates that the predicted distributions are not compatible with the
observed one at the 95.80 and 93.45 per cent level for the radio and X--ray
leading scenarios, respectively. 
A much more significant disagreement is found in the Slew survey 
(considering only XBLs), which is inconsistent with the predictions of
the radio leading scenario at the 99.96 per cent level. 

\subsection{Evolution} 

This last point on the redshift distribution is particularly important. 
In fact it indicates that evolutionary effects should be taken into
account. However it is possible that it is the distribution $\cal P$ which
changes with redshift, rather than the luminosity in one of the two bands: 
clearly one can re--express the different evolution in the two bands, by
saying that $\cal P$ evolves. 

From an observational point of view, if one considers different intervals
in redshifts in the Slew survey sample there are 5 RBL and 7 XBL objects
at high redshifts, while these numbers become 3 and 26 at low redshifts
(taking a dividing value at $z = 0.25$).  The only two XBL sources with
known redshift present in the 1 Jy sample are both at $z < 0.1$. 

We then considered an evolving probability distribution $\cal P$, i.e.  we
introduced a redshift dependence on the relative ratio of the two
populations of BL Lac objects.  Basically $\cal P$ has a bimodal shape,
described with two gaussians centered at $\log (L_{\rm 1keV}/L_{\rm 5GHz}) 
= -6.5$ and $-$3.5 for RBLs and XBLs, respectively. The relative
normalization of these gaussians evolves as $N_{\rm XBL}/N_{\rm RBL}
\propto (1 +
z)^{-\gamma}$ (with $\gamma \ge 0$), in such a way that the RBL part of
the distribution becomes more and more dominant at higher redshifts. With
this evolution of $\cal P$ we can induce the observed difference in the
redshift distribution of XBLs and RBLs.

In this case the luminosity function is not self--consistently derived.
Instead, we have to deal with a {\it tentative} luminosity function, in
the sense that it is by no means consistent with the evolution that we are
introducing. It is therefore necessary to derive ``backwards'' how the
evolution of $\cal P$ translates in terms of luminosity function
evolution.

We considered only a limited region of the parameter space of $\gamma$ and
$N_{\rm XBL}/N_{\rm RBL}|_{(z=0)}$, without finding a solution which
improves the
results already obtained. Actually, in the case of the radio leading
scenario, there is a hint that better solutions are obtained by changing
the parameters progressively towards the no evolution case, as in the G\&P
model.

While we cannot exclude that a suitable solution could be found following
these assumptions, we decided to consider a new approach ``a priori'',
described in the following section.

\section{Unified bolometric approach} 

\subsection{The idea} 

Partly motivated by the difficulties found with the models discussed so 
far, we tried a completely different way of facing the problem. 
Furthermore, we wanted to take into account the
properties of the overall observed SED of BL Lacs, as well as recent
indication of a possible link, along a continuous sequence, between the
SED shape and the source luminosity (e.g. Sambruna et al.  1996, 
Ghisellini et al., in preparation).

The fundamental
hypothesis of the model is still that XBL and RBL sources are 
manifestations of the same physical ph\ae nomenon. The new ingredient 
is the idea that XBLs and RBLs are
different representatives of a spectral sequence that can be described
in terms of a single parameter, which we identify with the {\it
bolometric luminosity of the synchrotron component}, $L_{\rm bol,sync}$.
Both ``flavours'' of BL Lac objects share the same bolometric luminosity
function and the SED properties depend {\it strongly} on this
quantity. 

The main positive feature of this approach is that it offers a more direct
interpretation in terms of the physical properties of these sources. In
fact:

\begin{description} 
\item[a)] the assumptions are largely independent of
the details of the observed statistical samples and moreover they are not
based on the choice of a leading spectral band; 

\item[b)] there is an immediate connection between the parameters of
solutions ``acceptable'' from a statistical point of view and the physical
conditions of the emitting plasma (see Section 7). 
\end{description}

\noindent 
The relation between the bolometric luminosity and the SED has
been based on observed trends.  More luminous objects seem to have
RBL--type spectral properties, with the peak of the energy distribution in
the IR--optical range and Compton dominated soft X--ray spectra. 

The less luminous sources tend to display XBL--like SEDs, steep soft
X--ray spectra and a synchrotron component peaking in the UV--soft X--ray
band (e.g. Fig.~1b). The distribution between this two extremes seems to
be continuous and we propose that it is governed by the now leading
parameter $L_{\rm bol,sync}$. Therefore, if we characterize the SED with
the frequency at which the (synchrotron) energy distribution has a
maximum, a fundamental (inverse) relation must exist between the
bolometric luminosity and this frequency. 
The inverse dependence is qualitatively based on the trend found by
Ghisellini et al. (in preparation).

Interestingly, note that the different redshift distributions of the two
kinds of BL Lacs could be a natural outcome of this scenario. XBLs
objects come from the lower (and richer) part of the luminosity function
and so they would dominate at low redshift, but they would disappear at
large distances despite the increase of the available volume. On the
contrary, RBLs, even though coming from the poorer part of the luminosity
function, would become predominant at higher redshifts, being still
detectable. 

In the next Sections we discuss a parameterization of the observed SEDs,
which allows us to quantify the predictions of this model. 

\subsection{SEDs parameterization} 

In order to reproduce the basic features of the observed SEDs we
considered a simple two component model. The synchrotron radio to soft
X--ray component is represented with a power law in the radio domain, with
spectral index $\alpha_s$, smoothly connecting, at $\nu_{\rm junct}$, with
a parabolic branch ranging up to $\nu_{\rm peak}$. Beyond $\nu_{\rm peak}$
the synchrotron component steepens parabolically.

The hard X--ray Compton component is simply represented with a single
power law with spectral index $\alpha_h$. The normalization at 1 keV of this
hard component ($L_{\rm 1keV,Comp}$) is kept fixed relative to the
synchrotron one at 5 GHz ($L_{\rm 5GHz}$). This immediately implies a
correlation between the radio and X--ray luminosity in sources where the
emission at 1 keV is dominated by the Compton emission.

The parameters describing this representation of the SED are then five: 
$\alpha_s$, $\alpha_h$, the hard (Compton) X--ray normalization K$=
\log(L_{\rm 1keV,Comp}/L_{\rm 5GHz})$, $\nu_{\rm junct}$ and either the
width of the parabolic branch $\sigma$ or $\nu_{\rm peak}$. This is because
we require that the parabolic branch matches the flat radio component at
$\nu_{\rm junct}$ also in its first derivative, and therefore we cannot fix
independently both its width and peak frequency. A schematic
representation of this parameterization is shown in Fig.~5, while its
analytical expression is reported in Appendix A. 

For simplicity, we use the luminosity at the peak of the synchrotron
component, $L_{\rm bol,sync}$ instead of the bolometric luminosity. In
fact, for various complete samples of BL Lacs, Sambruna et al. (1996) 
find that the ratio between $L_{\rm bol,sync}$ and $\nu_{\rm peak} L_{\rm
peak,sync}$ is consistent with being constant, $\simeq 8$.

We then specify the relation between $\nu_{\rm peak}$ (the frequency at
which the peak in the $\nu L_{\nu}$ representation occurs) and the value
of $\nu_{\rm peak} F'(\nu_{\rm peak})$ ($F'$ is the flux in the source
frame). This is the {\it key} physical relation of the proposed model. For
simplicity we considered a simple power law dependence $\nu_{\rm peak} =
\nu_{\rm peak,0} (L_{\rm peak}/L_{\rm peak,0})^{-\eta}$
($\eta > 0$)\footnote{We note that hereafter $L_{\rm peak}$ indicates the
equivalent of a bolometric luminosity, i.e. $\nu L_{\nu}$ evaluated at
$\nu_{\rm peak}$.}. 
We then have three more parameters: $\eta$, $L_{\rm peak,0}$ and $\nu_{\rm
peak,0}$. 

In Fig.~6 a set of SEDs following the adopted parameterization is shown. 
The dashed lines represent those with $\log (L_{\rm 1keV}/L_{\rm 5GHz}) >
-5.5$, roughly corresponding to XBL--type objects, while the opposite is
true for the solid ones, RBL--type objects. Remember that the distinction
is actually defined in terms of the flux ratio, and that $\log
(F_{\rm 1keV}/F_{\rm 5GHz}) = {\rm log}(L_{\rm 1keV}/L_{\rm 5GHz}) 
 + (\alpha_{R} - \alpha_X)\log(1 + z)$. 

To summarize: the adopted parameterization requires 8 inputs, 5 describing
the shape of the SED ($\alpha_s$, $\alpha_h$, $K$, $\nu_{\rm junct}$,
$\nu_{\rm peak}$) and three its relation with the luminosity
($\nu_{\rm peak,0}$,
$L_{\rm peak,0}$, $\eta$). However, the family of SEDs is completely
determined by a set of 7 parameters, because the first 5 reduce to 4 $+$
1, ($\alpha_s$, $\alpha_h$, $K$, $\nu_{\rm junct}$,
$\nu_{\rm peak}(\nu_{\rm peak,0},
L_{\rm peak,0}, \eta)$). In addition we must specify the normalization
$\Phi_0$ and the slope $\alpha_{\Phi}$ of the adopted luminosity function. 

\subsection{SEDs: parameterization vs observations} 

Despite the number of parameters, there is only a limited freedom in the
choice of their values, which are constrained by observations. 

\begin{description} \item[$-$] $\alpha_s$ only takes values in a narrow
range around 0, and we perform our simulations with the fiducial values
0.0, 0.1, 0.2. 

\item[$-$] The value of $\alpha_h$ is constrained by the assumption that,
in the case of extremely hard X--ray spectra, we are only looking at the
Compton component. We then considered a single value of $\alpha_h = 0.7$
equal to the slope of the flatter X--ray spectra of RBL (Lamer et al. 1996, 
Urry et al. 1996), for which we can think that we are measuring only 
their Compton emission. 

\item[$-$] On the same basis, the relative normalization between the radio
and the X--ray components can be constrained by the value of
$\log(L_{\rm 1keV}/L_{\rm 5GHz})$ of extreme RBL type objects, i.e. in the
range [$-$8 ; $-$7]. We adopted the values $K = -7.0$ and $K = -7.5$.

\item[$-$] $\eta$ can be limited by the ranges of observed $L_{\rm peak}$
and
$\nu_{\rm peak}$, which imply values of $\eta$ between $\sim $ 1 and 2.

\item[$-$] $\nu_{\rm junct}$ is the parameter less determined by
observations,
being difficult to identify a spectral break even in the case of the best
sampled and simultaneous SED.  We considered values between $10^{10}$ and
$10^{12}$ Hz.  In the context of this parameterization the actual
importance of $\nu_{\rm junct}$ is indirect: its position affects the
width of
the parabolic branch extending up to the X--ray band. For a fixed value of
$\nu_{\rm peak}$ the higher is $\nu_{\rm junct}$, the narrower will be the
parabola (i.e. the steeper the cut--off). \end{description}

Let us consider now some observed properties of complete BL Lac samples,
which allow us to check the goodness of our parametric representation. 

In Fig.~7, the classic $\alpha_{OX}$ vs $\alpha_{RO}$ plane for BL Lacs is
shown, with points representative of the 1 Jy, Slew and EMSS samples.
The line crossing the diagram is the locus of points 
derived from our ``best fit'' SEDs.
It basically reproduces the observed pattern of two different branches, 
where RBLs and  XBLs are concentrated, apparently with no overlapping.
The fact that the ``synthetic path'' does not extend beyond $\alpha_{OX}
\sim 0.8$ is related with the narrow range of parameters (and tight
relation among them) adopted in the samples simulation, namely to the
minimum allowed bolometric luminosity.

It is also possible to compare the SEDs with the observed average spectral
energy distributions of RBLs and XBLs, taken from Sambruna et al. (1996).
Their data points, given in flux units and converted in luminosities
(considering a typical redshift of 0.5 for the RBLs and 0.2 for XBLs), are
plotted in Fig.~6 together with our SED family. Also in this case the
agreement is satisfactory. 

Finally, also the $L_{\rm 1keV}$ vs $L_{\rm 5GHz}$ diagram (see Fig.~1b) 
can be interpreted in terms of the SEDs of RBLs and XBLs. There is a fixed
relation connecting the synchrotron radio luminosity and the inverse
Compton X--ray luminosity and this can be better observed in RBLs, where
more likely the Compton component dominates the X--ray emission.  RBLs
then occupy a locus following this correlation. The XBLs on the contrary
lie away from this line because their X--ray emission is progressively
dominated by a different component, the extreme part of the synchrotron
spectrum. 

We indeed verified that the luminosity correlation for RBL seen in Fig.~1b
is not purely due to the common dependence on redshift of $L_{\rm 1keV}$
and $L_{\rm 5GHz}$. A Kendall's $\tau$ test gives a correlation at
$>$99.99 per cent level, which is still present, even though at the 96.8
per cent level, when the redshift dependence is excluded.  Furthermore, we
tried to establish the real nature of the lack of observed sources in the
region around $\log L_{\rm 5GHz}$=33 and $\log L_{\rm 1keV}$=28.5. 
The results are shown in Fig.~1a, which is the equivalent plot, where fluxes
(and not luminosities) are considered. The vertical and oblique lines
represent the flux limit of the 1 Jy survey and the ratio of the fluxes
defining XBL and RBL (i.e. $-$5.5), respectively. 
The grey area of the diagram shows the position in this flux--flux plane of
putative sources with luminosities in the empty area of Fig.~1b (indicated 
with similar shading) when we imagine to move them in distance, covering
a range of redshift between $0 - 0.6$.  
It can be seen that there is no apparent observational bias against
detection of sources with those luminosities: this indeed suggests that
the lack of objects is plausibly linked to their intrinsic properties.
This fact seems therefore to qualitatively agree with the shape of the SED 
assumed in the bolometric scenario. 

Clearly, none of the above checks concerns the number density of objects. 
We adopted a luminosity function for $L_{\rm peak}$ inspired to that
calculated by Urry \& Padovani (1995) (see their Fig.~13) for the 1 Jy and
EMSS BL Lacs. Their (bivariate) luminosity function is obviously affected 
by selection effects both in the X--ray and radio bands, but constitutes 
the available distribution closest to a truly `bolometric' one. 
The normalization has been chosen in 
the range $\log\Phi(L_{\rm peak}=45.0) = -44.8$ to $-$44.5 and the index
$\alpha_{\Phi}
= -1.90$ to $-$2.3:  also on these parameters we have a limited freedom.
Note that the normalization is not actually an interesting parameter and,
as already mentioned in Section 3.3, varying $\Phi_0$ only affects the
absolute number of sources, neither the relative number of the two kinds
nor the average luminosities. Finally, we stress that for simplicity no 
evolution has been assumed in the bolometric scenario. 

To conclude this Section, we can say that our schematic representation of
the SEDs well reproduces the basic properties of the observed BL Lacs
broad band spectra.  We have 9 parameters and 14 observational quantities
to match (8 average luminosities and 6 source numbers).

\subsection{Results} 

Even a simple ``first attempt'' set of the 9 input parameters gives a
surprisingly good output. In Table~1 the best results, found after a
systematic check for a grid of parameter values within the allowed ranges,
are reported. The corresponding input parameters are listed in the note to
the Table. 

The Student's t--test implies that the model predicts too high 
luminosities for the RBL both in the radio and X--ray band 
(at the 99.9 and 98.6 per cent level, respectively, see Table~1) 
when compared with the observed 1 Jy Survey distributions.
The result is confirmed by the survival analysis (see Table~2).
For the Slew survey instead no significant disagreement has been found. 

Furthermore, a very positive consequence of this scenario is that it
implies a qualitatively `better' redshift distribution, as can be seen in
Figs.~3 and 4, with respect to the previous schemes. 
In particular we stress that the bolometric scenario is the only one which
predicts, as direct result of the SED dependence on the bolometric
luminosity, that the redshift distributions of the two populations can be 
such that RBL would tend to dominate at high 
redshifts, while the opposite would be true at low $z$ (see Fig.~4).
This behaviour could be seen as equivalent to either a redshift 
dependence of the X--ray-to-radio--flux ratio distribution $\cal P$
or a negative X--ray luminosity evolution. 

There is still a problem with RBLs in the radio survey. Even though the
shape of the $z$ distribution is correctly flat (and this is the major
difficulty of the two other models), it extends well beyond $z=1$, which
leads to a disagreement significant at the 92.7 per cent level 
(survival analysis). 
We believe this fact is directly connected with the excessive average 
radio luminosities predicted by the model. 

In order to explore the consistency of the {\it shape} of the predicted
and observed redshift distributions, we applied the survival analysis test 
only up to $z=1$, and did not find inconsistency between the two. 
with the observed one. Clearly, this result only indicates that the shape
of the predicted distribution can resemble the real one over the 
considered redshift interval. 
However, as we discuss in Section 7, the excess of
objects predicted at higher redshift (and/or their overestimated radio
power) could be understood in an even broader unifying picture. 

\section{Discussion and conclusions} 

We have considered three different scenarios, which assume that BL Lac
objects constitute a single population of sources, with different
spectral energy distributions (mainly different peak frequencies of the
synchrotron component), which originally mislead into thinking of a
separation of BL Lacs into two classes. 

A representation of the main features of these scenarios is shown in the
cartoons of Fig.~8a,b,c. 

Two of these models (namely the radio and X--ray leading ones; see
Fig.~8a,b) are symmetric, and differ in the choice of the leading
parameter, i.e. the radio and X--ray luminosities, respectively.  The
models are constructed from observed properties, like the luminosity
functions and the probability distribution $\cal P$ of the flux ratios,
with basically no free parameters. They can correctly predict a
significant number of quantities in agreement with current observations. 
However they (and in particular the radio leading one) fail to 
reproduce some of the observed distributions. 

We then proposed a unified bolometric model (see Fig.~8c), whose main
characteristic is to link the bolometric luminosity with the energy of the
synchrotron cut--off. This scenario is based too on observational trends,
however it uses a schematic parameterization of the BL Lac SED as well as
a semi--empirical luminosity function for the bolometric luminosities. We
stress here that while it is true that the bolometric scenario contains a
significant number of (quite constrained) parameters with respect to the
two other models, it also predicts the statistical and cosmological
properties of the two `types' of sources directly from a unified
description of their spectral distribution. And, despite the rigid
formulation of the one--to--one correspondence between the SED properties
and the luminosity, in this scenario the main observational data can be
reproduced at least with the same `accuracy' as the other two models.
Particularly interesting is the prediction of the different XBL and RBL
redshift distributions, which tend to favour the detection of objects of
the first class at low redshifts and of RBL at higher $z$. 

The predictions of the bolometric model of luminosities of sources detected 
in radio surveys higher than observed is plausibly related to
the problem with their redshift distribution. We note, in fact,
that the inclusion of Highly Polarized Quasars (HPQ)  (i.e. emission line
Blazars) in the $L_{\rm 1keV}$--$L_{\rm 5GHz}$ and colour--colour
distributions shows an interesting continuity of their properties with those
of BL Lacs (see Figs.~1 and 7, where stars represent HPQ).
In particular, the significance of the correlation between $L_{\rm 5GHz}$
and $L_{\rm 1keV}$ (see Section 6.3) increases to 99.99 per cent (already
excluding the common dependence on redshift) when also the HPQ of the Impey
\& Tapia (1990) sample are included. 
The morphology of the extended radio emission of some RBL (see
e.g. Kollgaard \etal 1992; Perlman \& Stocke 1993), and the radio luminosity
functions of RBLs and FSRQ (a class of Blazars substantially equivalent to
HPQ) also show continuity (Maraschi \& Rovetti 1994).  All these pieces of
evidence lead to the suggestion that there is actually a remarkable
progression in properties, in a luminosity sequence XBL--RBL--HPQ, which is
also a sequence of increasing importance of emission lines. If a fraction of
the sources with the highest bolometric luminosities could be indeed
``classified'' as HPQ rather than RBL (in a way analogous to the transition
from XBLs to RBLs), this fact could plausibly explain the discrepancies in
the high luminosities and redshift distributions predicted by the model.

As already stressed, a very interesting aspect of the ``bolometric''
approach is that it is based on the relative dependence of two quantities
which are strongly related to the physical properties of the emitting
plasma, namely the emitted luminosity and the cut--off in the synchrotron
spectrum.  One can therefore speculate on the physical origin of this
dependence. 

Probably the simplest interpretation of this link is that it is
directly related to the particle cooling. The more radiation is emitted
the more particles loose energy, with consequent decrease in the cut--off
frequency (which reflects the maximum energy of the emitting particles).
If the energy of the particles emitting at the peak of 
the flux distribution is determined by the escaping time from the 
emitting region, this can be estimated by equating the cooling and 
a typical escape timescales. This 
leads to the relation $\nu_{\rm peak} \propto (L_{\rm bol}/R)^{-2}$, where
$R$ is the size of the emitting region.  This dependence is 
reasonably  close to the one which predicts the best results of the model 
($\eta=1.5$). 
Clearly this interpretation is not unique and one can as easily envisage 
physical scenarios where a different relation between these two 
quantities is expected (e.g. in the case the source size is proportional to 
the central object mass). 

Furthermore the $\nu_{\rm peak}(L_{\rm bol,syn})$ relation can be also
attributed, in the context of inhomogeneous jet models, to a possible
dependence of the length of the collimated jet on the radiated power: more
powerful jets could extend farther and originate outer radio regions
responsible for the X--ray Compton component.

To conclude, we consider here the predictions of the models examined on the
results of deeper surveys. In particular we compute the expected ratio
of objects of the two ``flavours'' in more sensitive radio and X--ray
surveys, as a function of their flux limits. The results are reported in
Table~3\footnote{Browne \& March\~a (1993) pointed out the possible effect 
of misidentification of faint BL Lacs caused by the lack of contrast of 
the active nucleus with the host galaxy. It should be noted here that, while 
we were able to ignore
the Browne \& March\~a selection effect for the
results of the Slew survey, this would be progressively relevant in
surveys with decreasing flux limit, as the ones we are simulating
here.}, as the relative ratio $N_{\rm RBL}/N_{\rm XBL}$ normalized to the
value predicted for the 1 Jy and Slew surveys. 

Note that this ratio is a function of the flux limit even in the radio
survey for the radio leading scenario (and X--ray survey for the
X--ray leading one).  This is due to the different dominant range of
redshifts at different flux limits, which causes objects at the border 
of the
definition of XBL/RBL to ``move'' into the RBL class, because of the
term $(\alpha_R - \alpha_X)\log(1 + z)$ (always $<$ 0) in the relation
between flux and luminosity ratios.

Another factor to take into account in the interpretation of these
predictions is the influence of the minimum luminosities.  As mentioned in
Section 3.2, a minimum luminosity in the ``leading'' band leads to different
minimum luminosities for RBL and XBL in the ``secondary'' band, due to
their intrinsically different SEDs. Therefore, at flux limits which allow
the detection of the faintest sources, only objects with SEDs which favour
very low fluxes in that band will be detected. This effect is observable
mainly in radio surveys in the X--ray leading scenario and in X--ray
surveys in the radio leading one. 

From Table~3, one can clearly see the opposite trends expected from the
radio and X--ray leading scenarios in radio surveys.  Distinct features of
the bolometric scenario are a rapid decline in the fraction of RBL in
radio surveys and a weak decrease in X--ray surveys.  This behaviour
follows from the one--way relation between RBL and XBL SED properties with
luminosity.

\section*{Acknowledgments} 
We thank F. Rovetti and A.Treves for their helpful
collaboration in early stages of this work and the referee, Paolo 
Padovani, for his careful report and useful suggestions. 
GF and AC acknowledge the Italian MURST for financial support. 

\section{Appendix A: SED parameterization} 

The analytical expression of the parameterized SEDs in the
``bolometric'' model is given by: 

\begin{eqnarray*}
\psi_{s,1}(x)&=&\log(\nu F_{\nu})_{s,1}=  \\
         &=&\beta(x - 9.698) + \psi(5GHz) \\
\psi_{s,2}(x)&=&\log(\nu F_{\nu})_{s,2}=  \\
             &=&- \left({({x - x_{\rm peak})}/{\sigma}}\right)^2 +{0.25} \sigma^2\beta^2 + \psi_{s,1}(x)\\
\psi_{h}(x)&=&\log(\nu F_{\nu})_h= \\
           &=&(1 - \alpha_h)(x - 17.383) + \psi_{\rm Comp}(1keV) =\qquad\qquad\qquad \\
           &=&(1 - \alpha_h)(x - 17.383) + \psi(5GHz) + K + 7.685 \qquad ,
\\
\end{eqnarray*}
where $x~=~\log\nu$, and $\beta~=~(1~-~\alpha_s)$, and 
$\sigma = \left({2 ({x_{\rm peak} - x_{\rm junct})}/{\beta}}\right)^{1/2}$.
$\psi_{s,1}(x)$ and $\psi_{s,2}(x)$ are defined on the frequency ranges
$x < x_{\rm junct}$ and $x > x_{\rm junct}$, respectively.

The SED expression can be re-scaled as a function of the value of
$L_{\rm peak} = 10^{\psi(x_{\rm peak})}$, and in particular the
luminosities
at
5 GHz and 1 keV can then be written as: 
\begin{eqnarray*}
\psi(5GHz) & = & (1 + {0.5}\eta\beta)\psi_{\rm peak} -
{0.5}\beta(x_{\rm peak,0} + \\ 
& & +\eta\psi_{\rm peak,0} - x_{\rm junct}) - \beta(x_{\rm junct} -
9.698)\\ 
\psi_{\rm sync}(1keV) & = & \psi(x_{\rm peak}) - 
\left({{(17.383 - x_{\rm peak})}/{\sigma}}\right)^2 \\
\psi_{\rm Comp}(1keV) & = & \psi_{\rm sync}(5GHz) + K + 7.685 \\ 
\psi_{\rm tot}(1keV) & = & \log(10^{\psi_{\rm sync}} +
10^{\psi_{\rm Comp}})\\
\end{eqnarray*}

\section*{References} 

\refitem Blandford R., Rees M.J., 1978, in Pittsburgh Conference on 
BL Lac Objects, ed. A. N. Wolfe (Pittsburgh:Pittsburgh Univ. Press), 328

\refitem Browne I.W.A., March\~a M.J.M., 1993,  MNRAS, 261, 795

\refitem Brunner H., Lamer G., Worrall D.M., Staubert R., 1994, 
A\&A, 287, 436 

\refitem Celotti A., Maraschi L., Ghisellini G., Caccianiga A., 
Maccacaro T.,  1993, ApJ, 416, 118

\refitem Comastri A., Molendi S., Ghisellini G., 1995, MNRAS, 277, 297

\refitem Elvis M., Plummer D., Schachter J., Fabbiano G., 1992, ApJS, 80,
257

\refitem Feigelson E.D., Nelson P.I., 1985, ApJ, 293, 192

\refitem Fossati G., Maraschi L., Rovetti F., Treves A., 1996, Proc. of the
Conference ``R\"ontgenstrahlung from the Universe'', W\"urzburg September
1995, MPE Report 263, 447

\refitem Ghisellini G., Maraschi L., 1989, ApJ, 340, 181  

\refitem Giommi P., Ansari S.G., Micol A., 1995, A\&AS, 109, 267

\refitem Giommi P., Padovani P., 1994, MNRAS, 268, L51 

\refitem Impey C.D., Tapia S., 1990, ApJ, 354, 124

\refitem Jannuzi B. T., Smith P. S., Elston R., 1994, ApJ, 428, 130

\refitem Kollgaard R.I., Wardle J.F.C., Roberts D.H., Gabudza D.C., 1992, AJ, 104, 1687 

\refitem Kollgaard R.I., 1994, Vistas in Astronomy, 38, 29

\refitem Kollgaard R.I., Palma C., Laurent--Muehleisen S.A., Feigelson E.D., 1996, ApJ, 465, 115

\refitem K\"uhr H., Witzel A., Pauliny--Toth I.K., Nauber U., 1981, 
A\&A, 45, 367


\refitem Lamer G., Brunner H., Staubert R., 1996, A\&A, 311, 384

\refitem La Valley M., Isobe T., Feigelson E.D., ``ASURV'' B.A.A.S. 1992

\refitem Maraschi L., Ghisellini G., Tanzi E., Treves A., 1986, 
ApJ, 310, 325

\refitem Maraschi L., Rovetti F., 1994, ApJ, 436, 79

\refitem Maraschi L., Fossati G., Tagliaferri G., Treves A., 1995, 
ApJ, 443, 578

\refitem Morris S.L., Stocke J.T., Gioia I.M., Schild R.E., 
Wolter A., Maccacaro T., Della Ceca R.,  1991, ApJ, 380, 49

\refitem Padovani P., Giommi P., 1995, ApJ, 444, 567 


\refitem Padovani P., Giommi P., 1996, MNRAS, 279, 526 

\refitem Perlman E.S., Stocke J.T., 1993, ApJ, 406, 430 

\refitem Perlman E.S., \etal, 1996a, ApJS, 104, 251  

\refitem Perlman E.S., Stocke J.T., Wang Q.D., Morris S.L., 1996b, 
ApJ, 456, 451 

\refitem Sambruna R., Maraschi L., Urry C.M., 1996, ApJ, 463, 444

\refitem Stickel M., Padovani P., Urry C.M., Fried J.W., K\"uhr H., 1991,
ApJ, 374, 431

\refitem Stickel M., K\"uhr H., 1994, A\&AS, 103, 349

\refitem Stickel M., K\"uhr H., 1996, A\&AS, 115, 1

\refitem Stickel M., Meisenheimer K., K\"uhr H., 1994, A\&AS, 105, 211

\refitem Stocke J.T., Liebert J., Schmidt G., Gioia I., Maccacaro T., 
Schild R., Maccagni D., Arp H., 1985, ApJ, 298, 619

\refitem Urry C.M., Padovani P., Stickel M., 1991, ApJ, 382, 501

\refitem Urry C.M., Padovani P., 1995, PASP, 107, 803

\refitem Urry C.M., Sambruna R., Worrall D.M., Kollgaard R.I., 
Feigelson E.D., Perlman E.S., Stocke T.S., 1996, ApJ, 463, 424

\refitem von Montigny C., et al., 1995, ApJ, 440, 525

\refitem Wolter A., Caccianiga A., Della Ceca R.,  Maccacaro T., 1994,
ApJ, 433, 29

\refitem Wurtz, 1994, PhD Thesis, Univ. Colorado

\vfill\eject
\clearpage


\begin{figure*}
\centerline{\psfig{figure=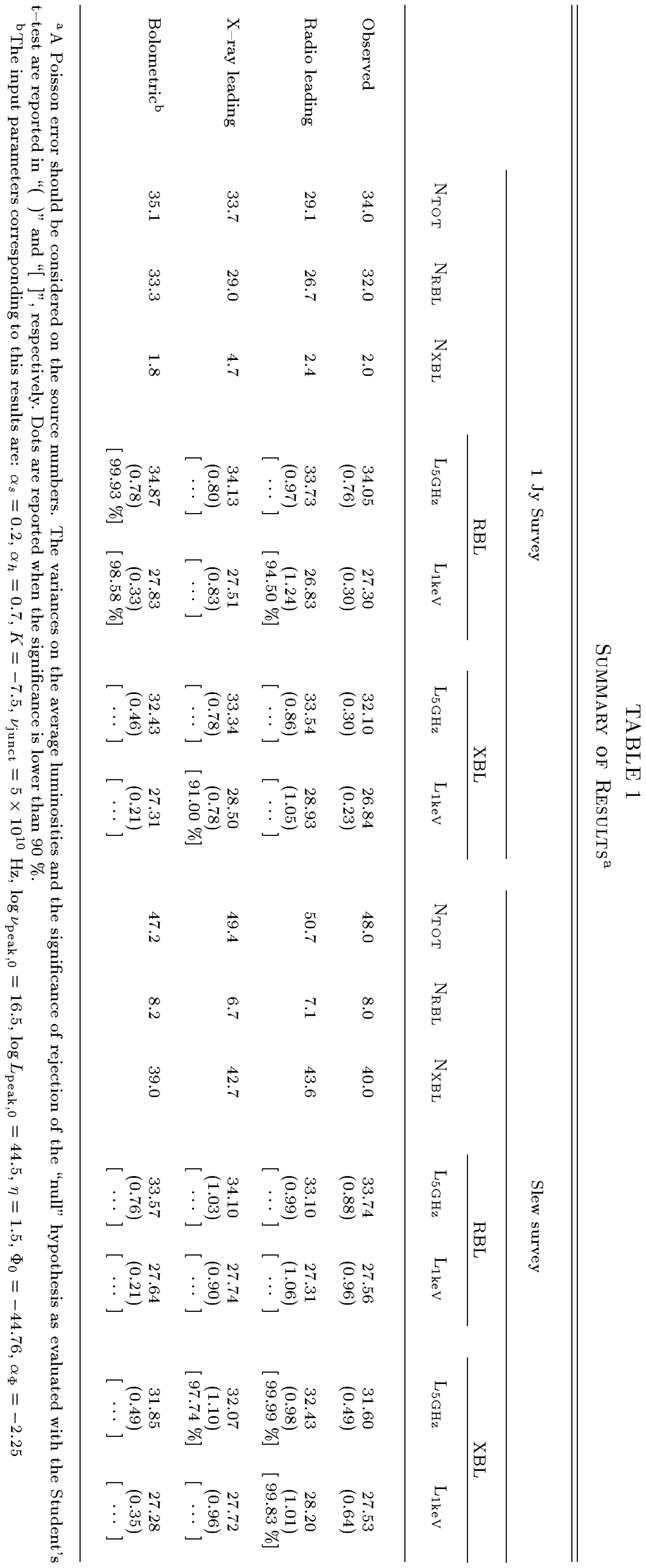,height=27truecm}}
\end{figure*}

\begin{figure*}
\centerline{\psfig{figure=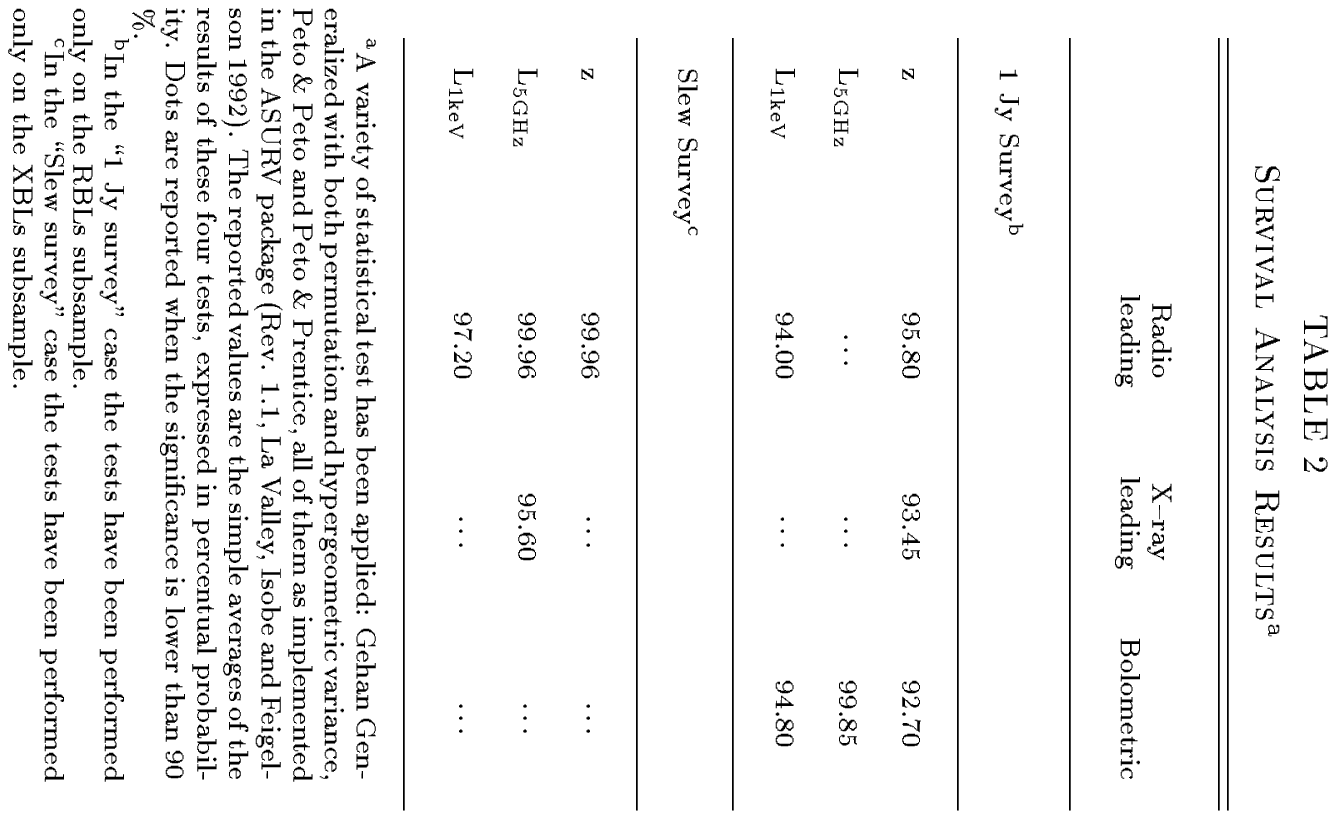}}
\end{figure*}

\begin{figure*}
\centerline{\psfig{figure=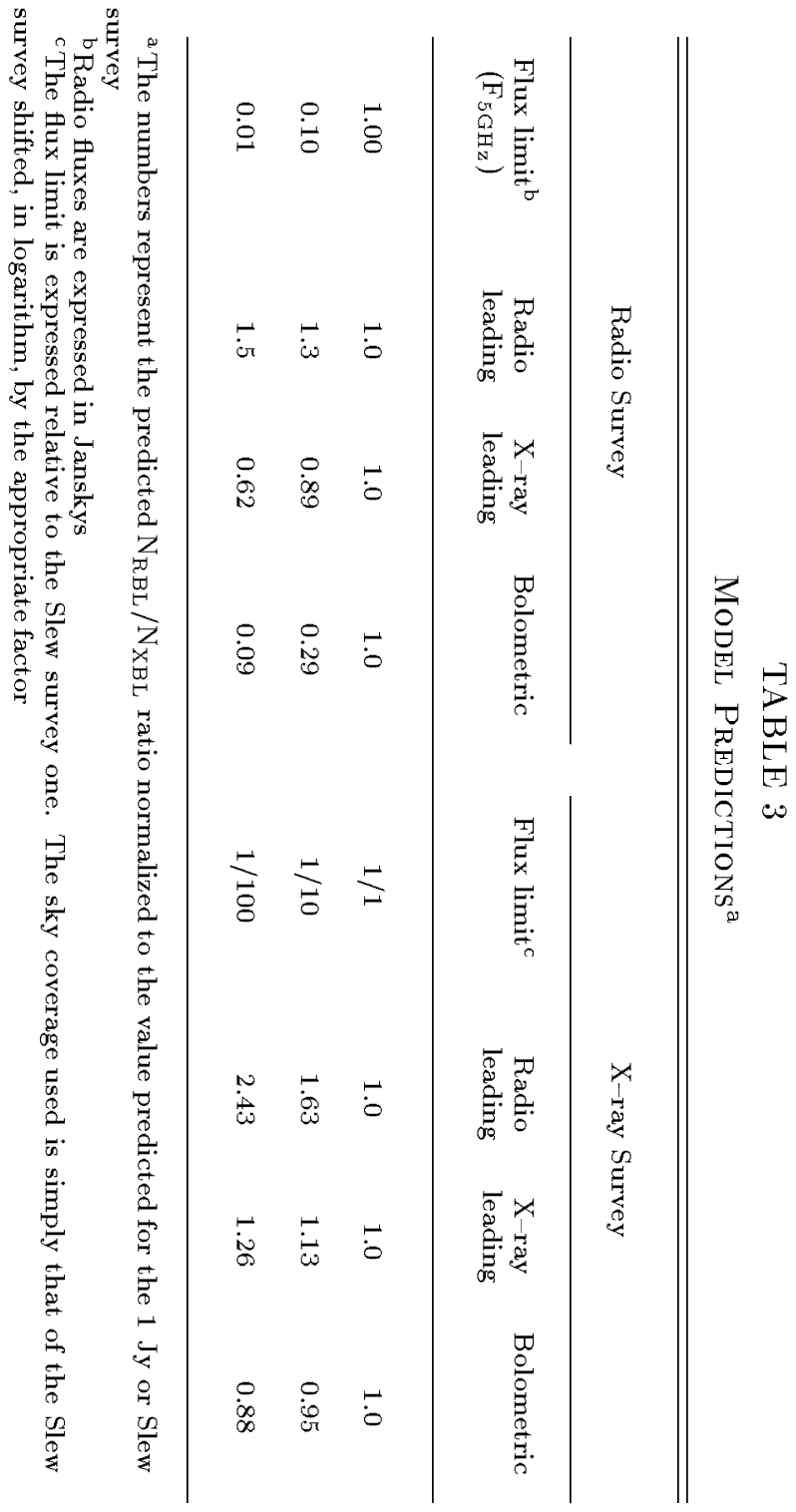}}
\end{figure*}

\begin{figure*}
\centerline{\psfig{figure=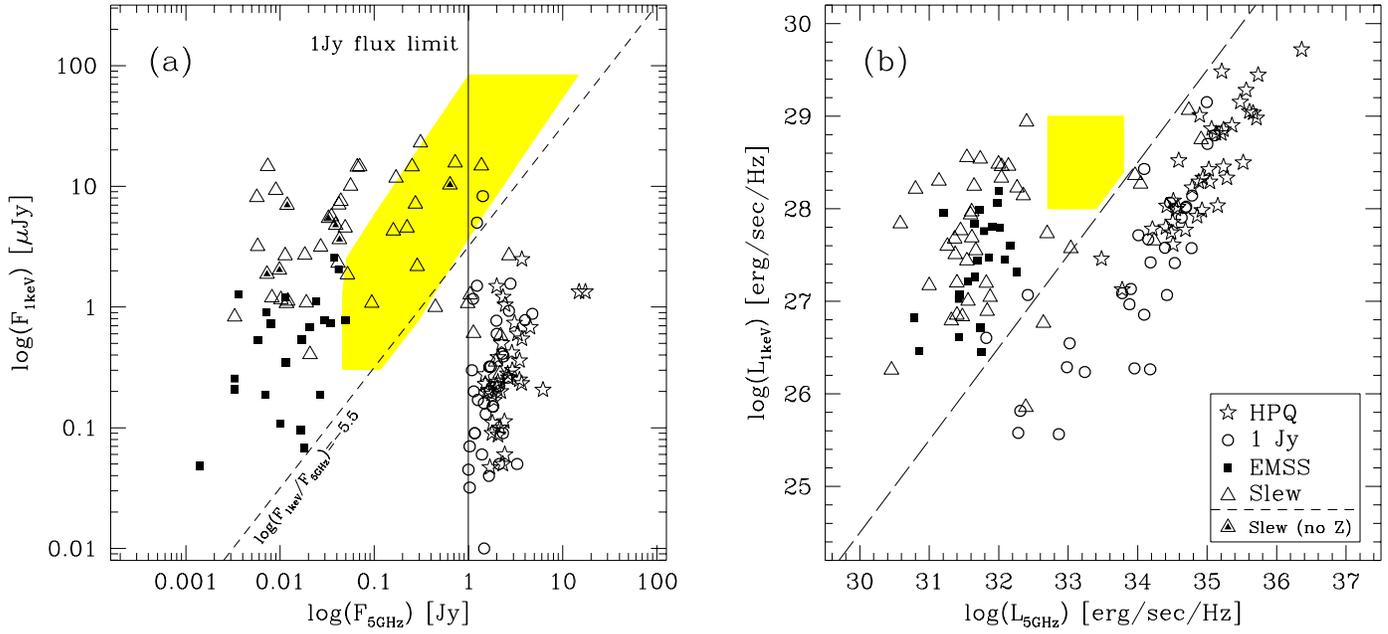,rheight=13truecm}}
\caption{(a) The $\log F_{\rm 1keV}$ vs $\log
F_{\rm 5GHz}$ plane for different samples: 1 Jy, EMSS, Slew survey, and
the complete HPQ sample from Impey \& Tapia (1990) (all the corresponding
symbols are defined in the inlet in panel (b)). The vertical line represents
the 1Jy flux detection limit, while the dashed line defines the
separation between XBL and RBL. Finally, grey area shows the loci of
points that potential sources shown in Fig.~1b in the similar grey region, 
would occupy in a $F_{\rm 1keV}$ vs $F_{\rm 5GHz}$ diagram for different 
redshifts ($0 < z < 0.6$). 
This shows that no obvious selection effects would avoid the
detection of these sources. (b) Distribution of luminosities for the same
source samples. The oblique line indicates the values of
$L_{\rm 1keV}/L_{\rm 5GHz}$ roughly corresponding to the definition of RBL
and XBL. }
\end{figure*}

\clearpage

\begin{figure*}
\centerline{\psfig{figure=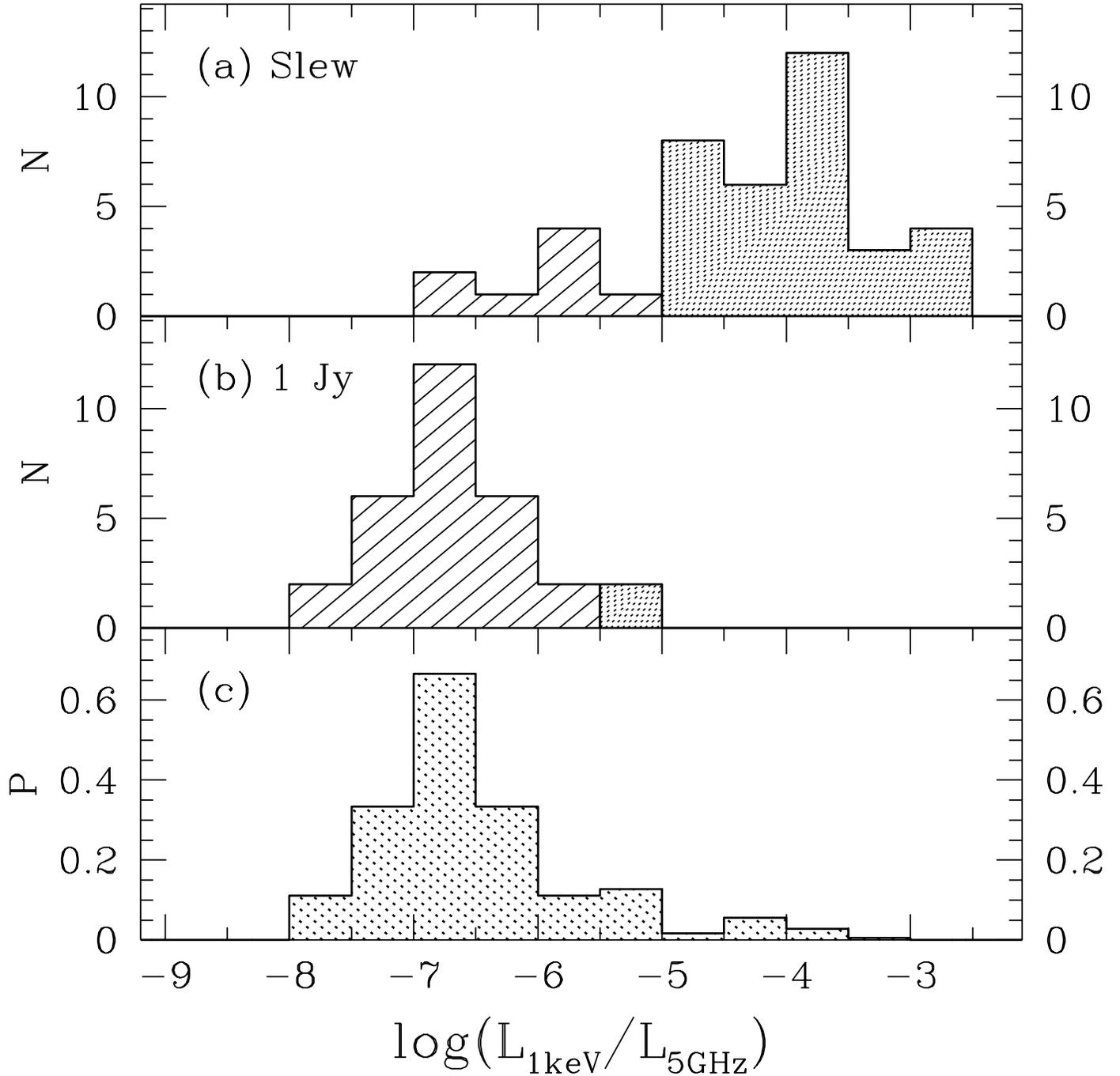}}
\caption{(a,b) Distribution of the
observed ratio $L_{\rm 1keV}/L_{\rm 5GHz}$ for the two surveys. Dashed
areas represent RBL, while dotted areas indicate XBL sources. The
distribution ${\cal P}(L_{\rm 1keV}/L_{\rm 5GHz})$ used in the
computations of the radio leading scenario is shown in (c). It is obtained
summing up that observed for the RBLs in the 1Jy sample and that of the
EMSS sample, weighted by a factor 1/10.}
\end{figure*}

\begin{figure*}
\centerline{\psfig{figure=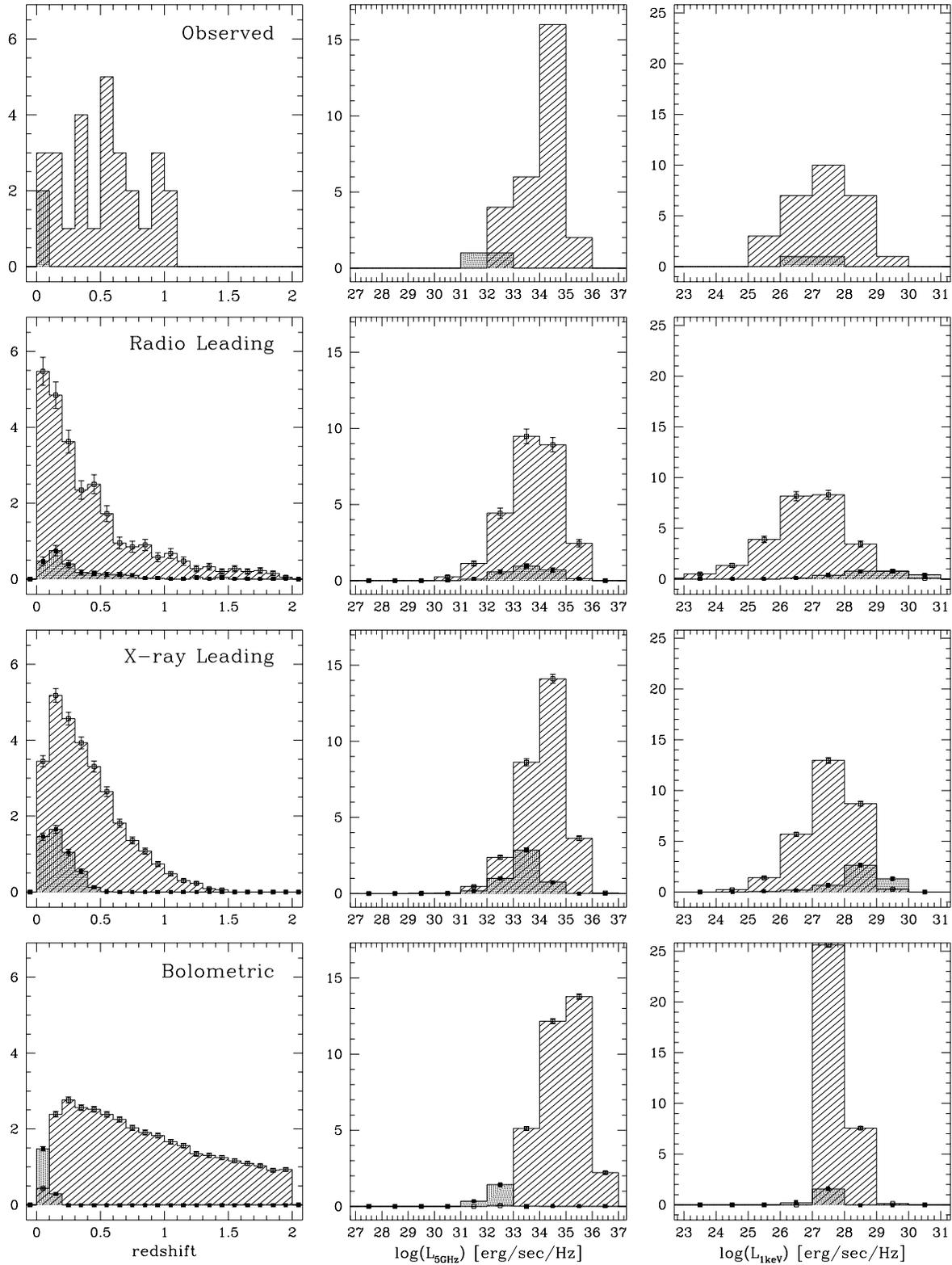,height=22truecm,rheight=22truecm}}
\caption{Results of the simulations of the
1 Jy survey compared with observations. From top to bottom:  observations,
``radio leading'', ``X--ray leading'' and ``bolometric'' models,
respectively. From left to right: histograms of the redshift
distributions, radio and X--ray luminosities for the two classes. Again
dashed and dotted areas represent RBL and XBL, respectively. The error
bars are derived from the simulations. }
\end{figure*}

\begin{figure*}
\centerline{\psfig{figure=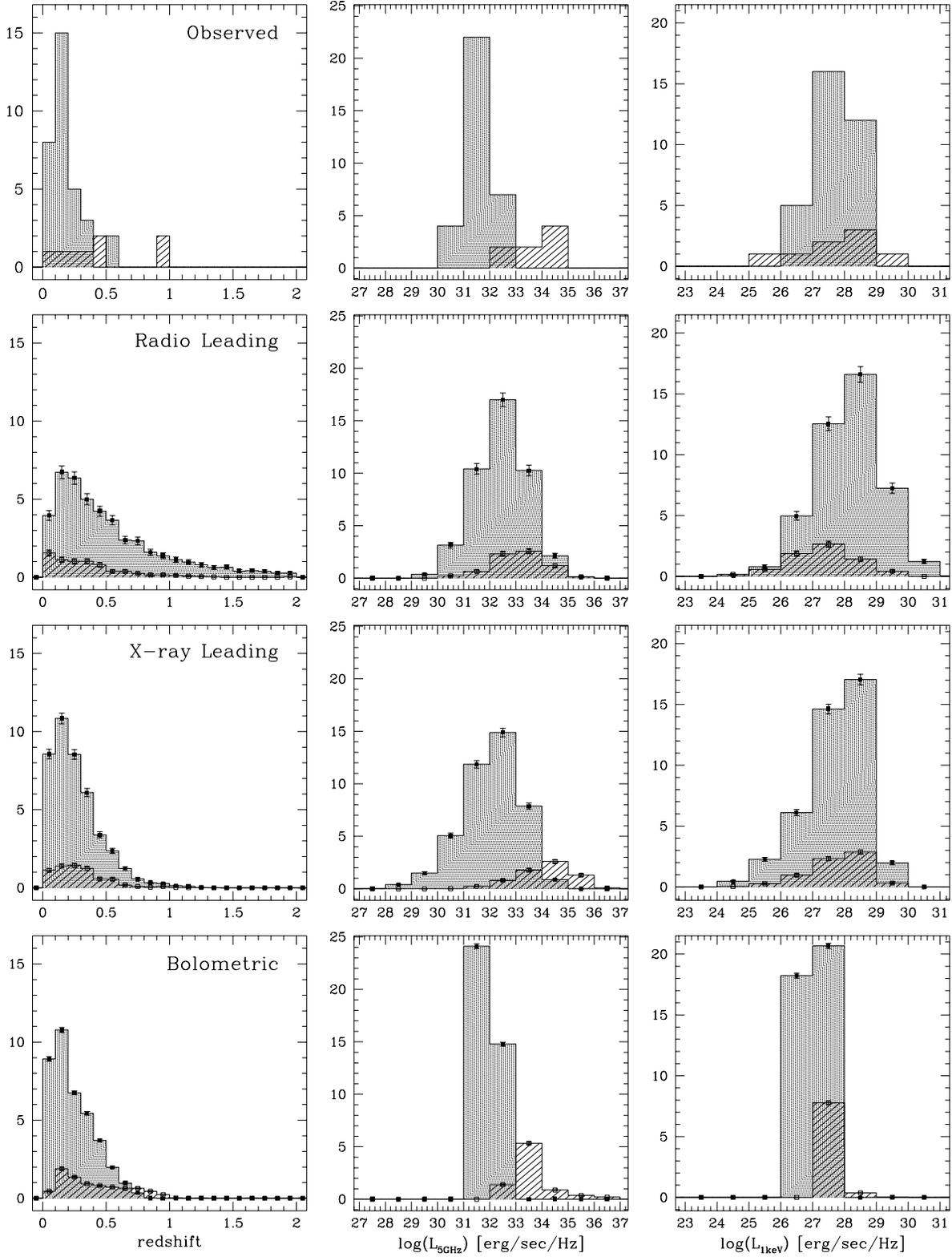,height=22truecm,rheight=22truecm}}
\caption{ Same as Fig.~3 for the simulated Slew survey. }
\end{figure*}

\begin{figure*}
\centerline{\psfig{figure=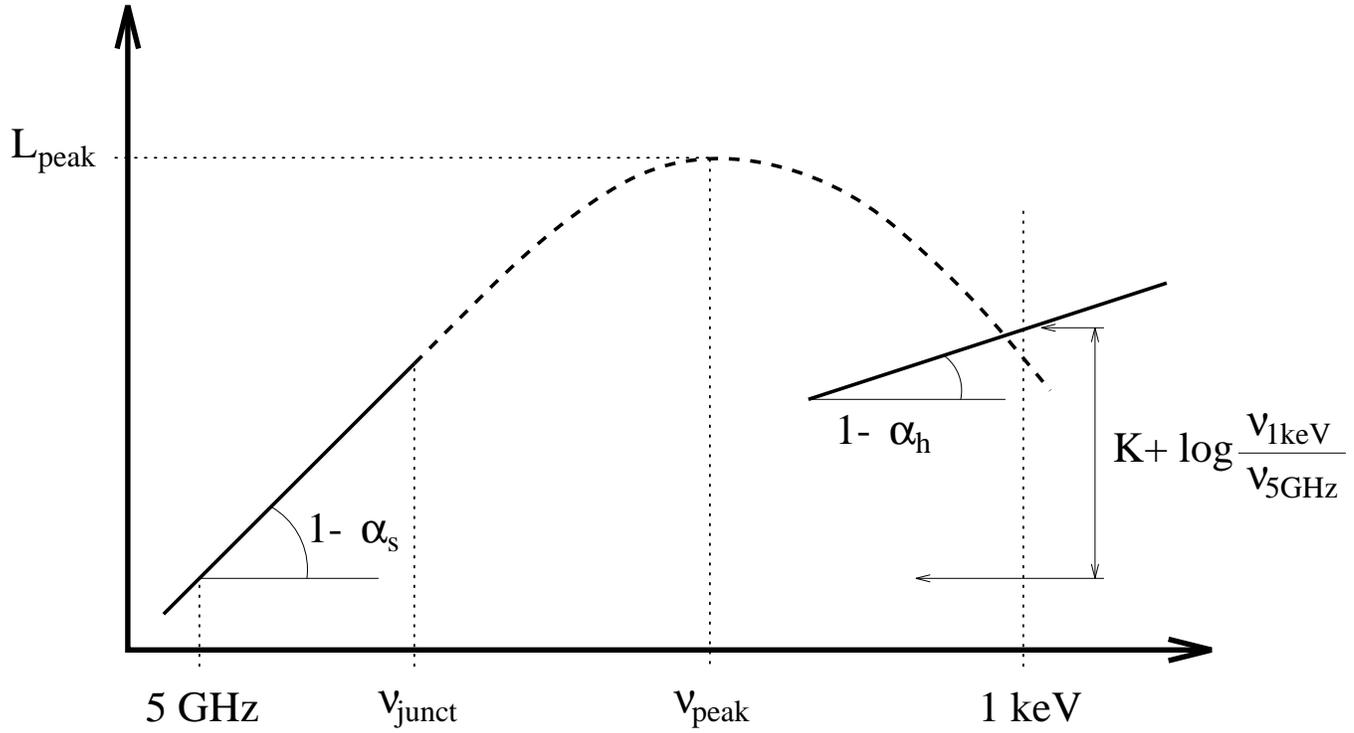}}
\caption{Schematic representation of the
parameterization of the SEDs, according to the ``unified bolometric''  
model. All the basic parameters are shown. Note that the difference
between radio and X--ray luminosities in the $\nu L_{\nu}$ representation
differs from the relative normalization of the two component, defined as
K, by the term $\log(\nu_{\rm 1keV}/\nu_{\rm 5GHz})$.}
\end{figure*}

\begin{figure*}
\centerline{\psfig{figure=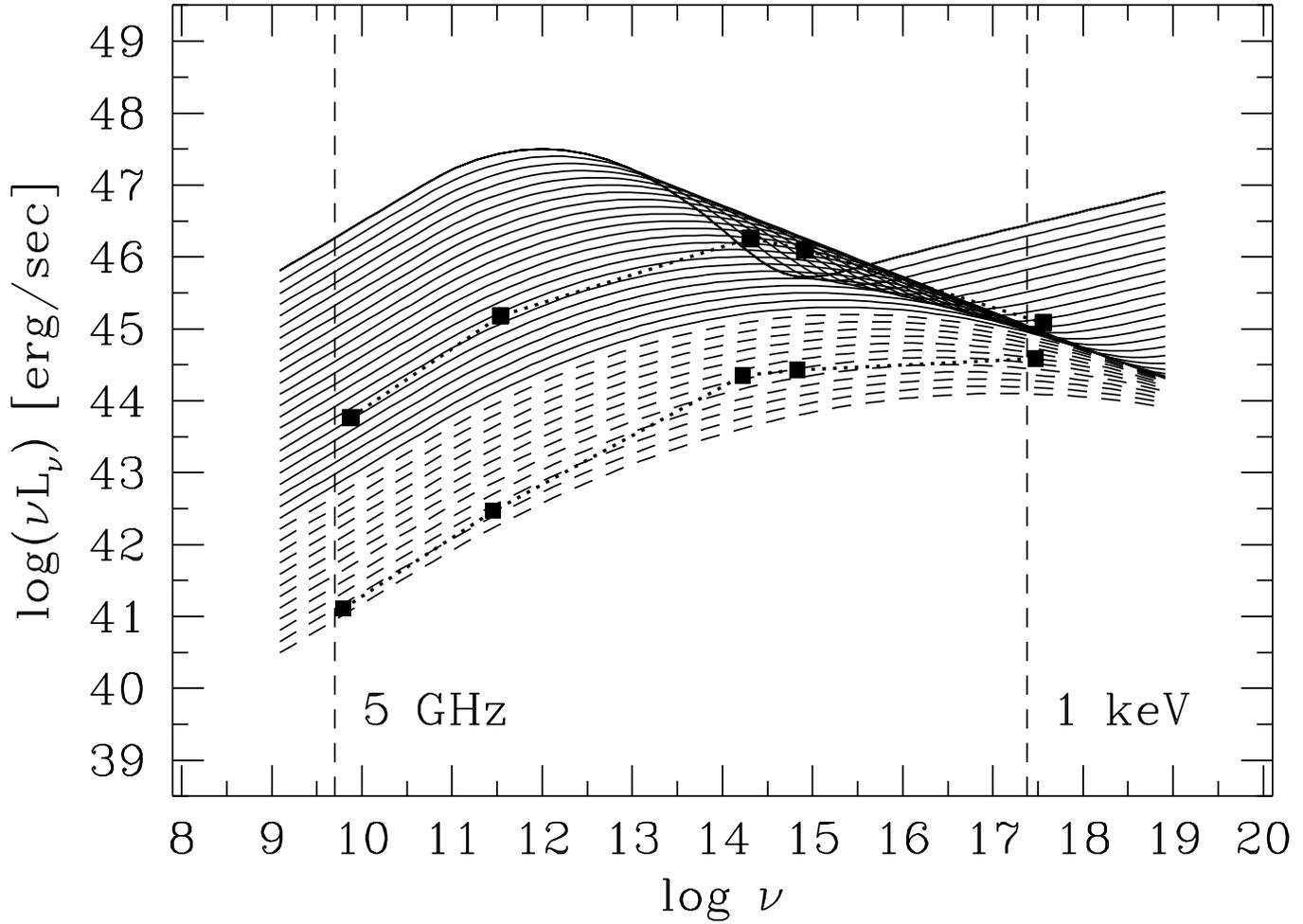,rheight=14.5truecm}}
\caption{Set of SEDs used in the simulations of the ``unified bolometric''
scenario. The dashed SEDs are those with $\log (L_{\rm 1keV}/L_{\rm 5GHz}) >
-5.5$, corresponding to XBL objects.  The solid SEDs, on the contrary,
correspond to RBL objects. Note that this subdivision is not strictly
equivalent to the classification defined in terms of the flux ratio, that
differs by a term $[(\alpha_{R} - \alpha_X) \log(1+z)] \simeq -0.3$. The
squares indicate the average observed SEDs of XBL and RBL, as reported by
Sambruna et al. (1996).}
\end{figure*}

\begin{figure*}
\centerline{\psfig{figure=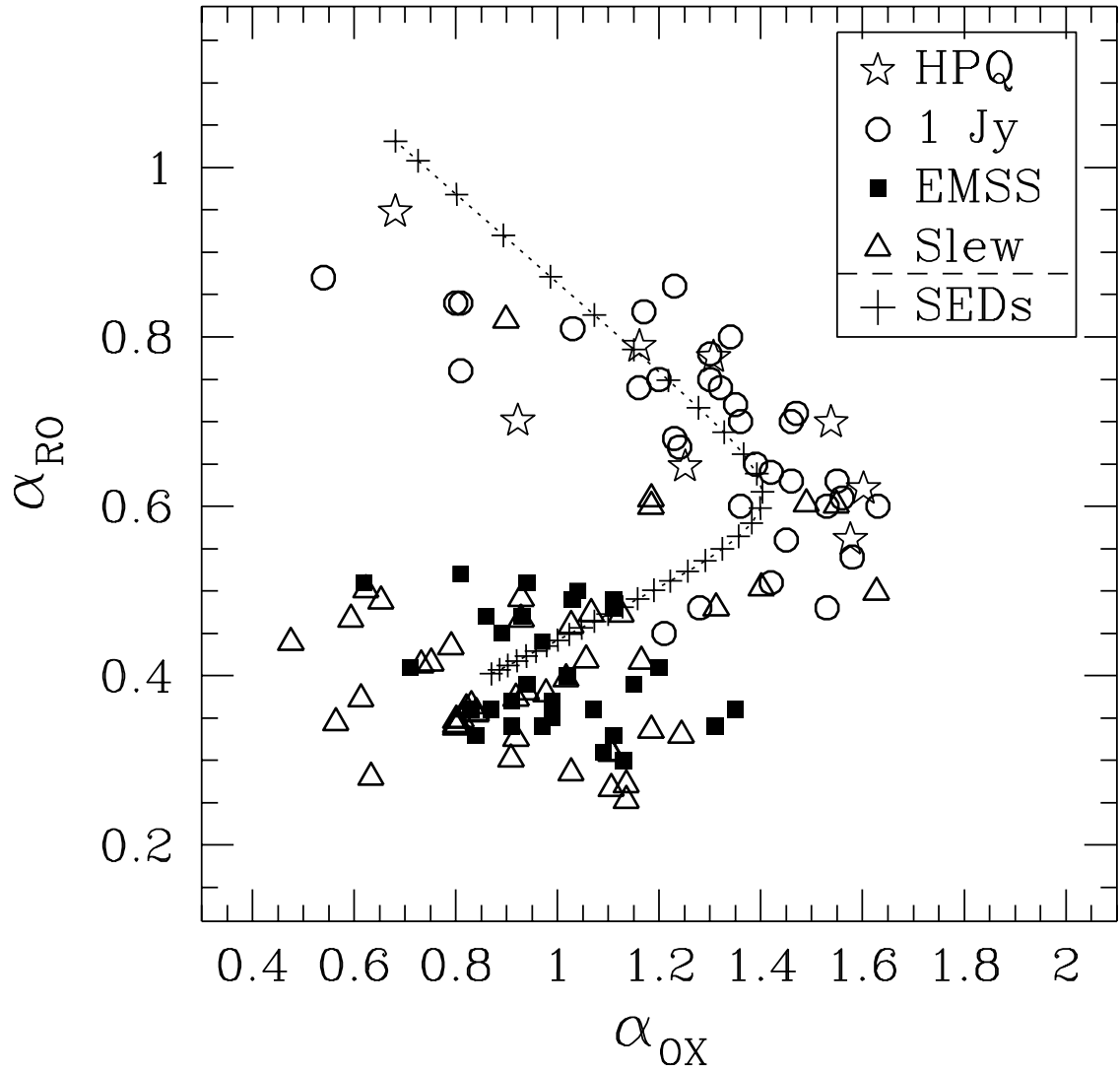,rheight=18truecm}}
\caption{Distribution of sources in the colour--colour diagram, i.e. in the
$\alpha_{OX}$ vs $\alpha_{RO}$ plane. 
Different complete samples are considered: EMSS, Slew and 1 Jy BL Lacs,
and High Polarization Quasars (HPQ) from the Brunner et al. (1996) sample.
All the corresponding symbols are defined in the inlet. 
The crosses connected by a dotted line show the spectral indices predicted 
by the adopted ``best fit'' parameterization of the SEDs. }
\end{figure*}

\begin{figure*}
\centerline{\psfig{figure=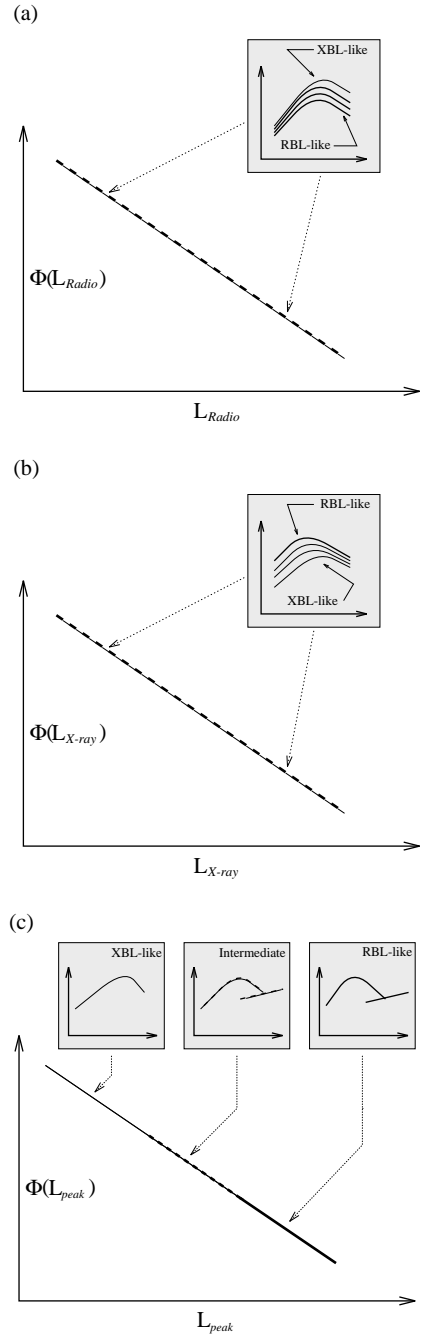,height=24truecm,rheight=20truecm}}
\caption{Cartoons resuming the basic features and differences between (a)
the radio leading, (b) the X--ray leading and (c) the new unified bolometric
scenarios. The plane defined by the two axis shows the leading luminosity
function. As shown in the panels in the first two schemes the SED
distributions and populations ratios are independent from luminosity, while
the bolometric approach links the type of source (i.e.  the SED) with the
luminosity. The schematic SED represented with thick and thin lines always
refer to RBL and XBL, respectively, and the number of SEDs in each panel is
suggestive of the relative number density of sources.}
\end{figure*}

\end{document}